\documentclass[12pt,preprint]{aastex}  

\begin{document}

\title{The Size Distribution of the Neptune Trojans and the Missing Intermediate Sized Planetesimals}
\author{Scott S. Sheppard}    
\affil{Department of  Terrestrial Magnetism, Carnegie Institution of Washington, \\
5241 Broad Branch Rd. NW, Washington, DC 20015 \\ sheppard@dtm.ciw.edu}

\and

\author{Chadwick A. Trujillo}
\affil{Gemini Observatory \\
670 North A`ohoku Place, Hilo, HI 96720}

\begin{abstract}  

We present an ultra-deep survey for Neptune Trojans using the Subaru
8.2-m and Magellan 6.5-m telescopes.  The survey reached a $50\%$
detection efficiency in the R-band at $m_{R}=25.7$ magnitudes and
covered 49 square degrees of sky.  $m_{R}=25.7$ mags corresponds to
Neptune Trojans that are about 16 km in radius (assuming an albedo of
0.05).  A paucity of smaller Neptune Trojans (radii $<$ 45 km)
compared to larger ones was found.  The brightest Neptune Trojans
appear to follow a steep power-law slope ($q = 5\pm1$) similar to the
brightest objects in the other known stable reservoirs such as the
Kuiper Belt, Jupiter Trojans and main belt asteroids.  We find a
roll-over for the Neptune Trojans that occurs around a radii of
$r=45\pm10$ km ($m_{R}=23.5\pm0.3$), which is also very similar to the
other stable reservoirs.  All the observed stable regions in the the
solar system show evidence for Missing Intermediate Sized
Planetesimals (MISPs).  This indicates a primordial and not
collisional origin, which suggests planetesimal formation proceeded
directly from small to large objects.  The scarcity of intermediate
and smaller sized Neptune Trojans may limit them as being a strong
source for the short period comets.

\end{abstract}


\section{Introduction}

Our understanding of how objects in the early solar system coagulated
to form planetsesimals of the kilometer size scale is mostly limited
to theory (Bottke et al. 2005; Johansen et al. 2007; Blum \& Wurm
2008; Cuzzi et al. 2008; Morbidelli et al. 2009).  Though we can
detect some gas and dust disks as well as large planets around stars,
we will not be able to detect extra-solar planetesimals on the
kilometer to thousands of kilometer size scale in the foreseeable
future.  Currently, the only way to directly study such a population
is through the stable reservoirs in our solar system (Jewitt et
al. 2000; Kenyon \& Bromley 2004; Pan \& Sari 2005; Bottke et
al. 2005; Fraser \& Kavelaars 2008; Fuentes \& Holman 2008; Fraser et
al. 2008; Fuentes et al. 2009; Fraser 2009; Morbidelli et al. 2009).
The orbits of objects in the main asteroid belt, Kuiper belt and
Trojan regions have been highly influenced by the migration and
evolution of the Solar System.  Since Trojan asteroids share their
planet's orbital period and semi-major axis they are especially useful
in constraining the formation and migration of their planet
(Morbidelli et al. 2005; Tsiganis et al. 2005).  Trojan asteroids lead
(L4) or trail (L5) a planet by about sixty degrees near the two
triangular Lagrangian points of gravitational stability.  The Jupiter
Trojans have been known since Max Wolf discovered 588 Achilles in
1906.  There are currently about 3000 known Trojans in the L4 and L5
regions of Jupiter.  Neptune's first Trojan was discovered within the
L4 region in 2001 while several more Neptune Trojans have been
discovered in recent years (Chiang et al. 2003; Sheppard \& Trujillo
2006, 2010; Becker et al. 2008).  The other giant planets Saturn and
Uranus are not expected to have a significant number of Trojans since
their Lagrangian regions are more dynamically unstable over the age of
the solar system (Nesvorny \& Dones 2002).

The dynamics of Kuiper Belt objects in outer mean-motion resonances
with Neptune, such as the 3:2 and 2:1 resonances, suggest that Neptune
likely migrated several AU outwards during its lifetime (Hahn \&
Malhotra 2005; Chiang \& Lithwick 2005; Levison et al. 2008).
Similarly, the Hildas in the main asteroid belt are in the 2:3
mean-motion resonance with Jupiter and their orbital distribution
suggests Jupiter migrated inward by a few tenths of AU (Franklin et
al. 2004).  Unlike these inner and outer resonance objects, the
Trojans, which are in a 1:1 resonance with their respective planet,
would likely have been depleted during any large, irregular planetary
migration.  The capture of Trojans in the current Solar System is not
an effective process and thus capture happened when the Solar System
dynamics were significantly different than now (Horner \& Evans
2006). Thus the Trojans were likely captured during either a slow
smooth planetary migration process or more likely after any
significant planetary migration through a Neptune freeze-in
circularization process (Kortenkamp et al. 2004; Morbidelli et
al. 2005; Sheppard \& Trujillo 2006; Li et al. 2007; Nesvorny \&
Vokrouhlicky 2009; Lykawka et al. 2009).  Both the L4 and L5 Neptune
Trojan regions appear to have similar sized populations and dynamics.
The expected large number of high inclination Neptune Trojans in both
the L4 and L5 regions of Neptune suggests that capture occurred with a
dynamically excited planetesimal population (Sheppard \& Trujillo
2010).

\section{Observations}

Observations were obtained in June 2008 and 2009 for the L5 region and
October 2004-2009 for the L4 region of Neptune.  For the Subaru
observations the Suprime-Cam with ten $2048 \times 4096$ pixel CCDs
arranged in a $5 \times 2$ pattern was used.  Suprime-Cam has $15
\micron$ pixels that gives a pixel scale of $0.\arcsec 20$
pixel$^{-1}$ at prime focus and a field-of-view that is about
$34\arcmin \times 27\arcmin$ with the North-South direction aligned
with the long axis.  Gaps between the chips are about $16 \arcsec$ in
the North-South direction and $3 \arcsec$ in the East-West direction
(Miyazaki et al. 2002).  At Magellan the the IMACS camera on the Baade
telescope was used.  IMACS is a wide-field CCD imager that has eight
$2048\times4096$ pixel CCDs with a pixel scale of $0.\arcsec 20$
pixel$^{-1}$ and a field-of-view of about 0.2 square degrees.  The
eight CCDs are arranged in a $4 \times 2$ box pattern with four above
and four below and about 12 arcsecond gaps between chips.

The images were bias-subtracted and then flat-fielded with twilight
flats.  During exposures the telescope was autoguided sidereally on
field stars.  Integration times were between 300 and 450 seconds and
images of the same field were obtained on three visits with 1 to 1.5
hours between visits.  Observations were obtained while the Neptune
Trojans were within 1 hour of opposition so the dominant apparent
motion was largely parallactic and thus is inversely related to
distance.  Objects at the heliocentric distance of Neptune, $R\sim 30$
AU, will have an apparent motion of about $\sim 4\arcsec $~hr$^{-1}$.

\section{Results and Analysis}

Previous surveys for Neptune Trojans focused mostly on finding
relatively bright objects (radii greater than 40 km or brighter than
24th magnitude) (Chiang et al. 2003; Sheppard \& Trujillo 2006; Becker
et al. 2008).  The Neptune Trojan regions cover thousands of square
degrees on the sky as many Trojans have large inclinations and librate
up to $\pm 30$ degrees from the Lagrangian points over a ten thousand
year time scale (Chiang et al. 2003).  We obtained an ultra-deep,
large area survey of the Neptune L5 and L4 Trojan regions.  A total of
49 square degrees were searched of which 30 square degrees were in the
L4 region while the other 19 square degrees were in the L5 region of
Neptune.  Subaru was used to covered 21 square degrees while Magellan
was used for the other 28 square degrees of the survey.  The first
high inclination Neptune Trojan, 2005 TN$_{53}$, was discovered at
Magellan (Sheppard \& Trujillo 2006) while the first L5 Neptune Trojan
(2008 LC$_{18}$: Sheppard \& Trujillo 2010) was found at Subaru.  In
all, five L4 Neptune Trojans were detected at Magellan ranging from
22.5 to 23.7 magnitudes (about 70 to 40 km in radius) while one L5
Neptune Trojan was detected at Subaru with an magnitude of 23.2 in the
R-band.

The data were analyzed with a computer algorithm tuned to detect
objects which appeared in all three images from one night with an
apparent motion consistent with being beyond the orbit of Jupiter
(speeds less than 20 arcseconds per hour).  Objects were flagged as
possible Neptune Trojans if they moved between 3.5 and 4.5 arcseconds
per hour.  These objects were recovered up to two months later to
determine if they had Neptune Trojan like orbits.  The survey was
designed similar to our ultra-deep surveys for satellites around the
planets (Sheppard et al. 2005; Sheppard \& Trujillo 2009).

\begin{figure}
\epsscale{0.4}
\centerline{\includegraphics[angle=90,width=\textwidth]{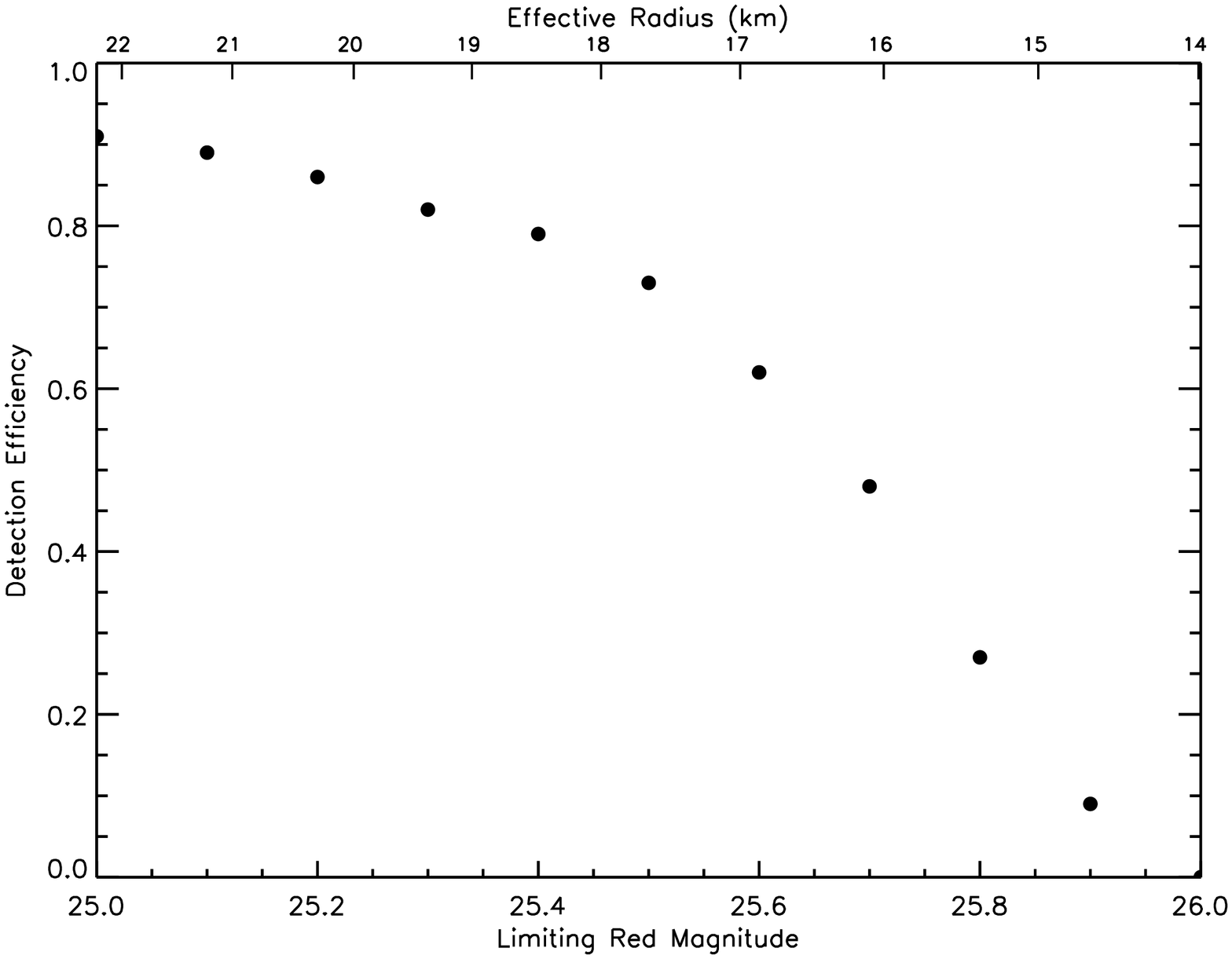}}
\caption{Detection efficiency of the Neptune Trojan survey versus
apparent red magnitude.  The 50\% detection efficiency is at an R-band
of 25.7 magnitudes as determined by our computer algorithm's
performance on implanted artificial moving objects at Neptune Trojan
apparent velocities.  Effective radii of the apparent magnitude were
calculated assuming an albedo of 0.05.}
\label{fig:effTrojans}
\end{figure}

We determined the limiting magnitude of the survey by placing
artificial objects in the fields matched to the point spread function
of the images and with motions of 4 arcseconds per hour.  The
brightnesses of the objects were binned by 0.1 mag and spanned the
range from 25 to 27 magnitudes.  Results are shown in
Figure~\ref{fig:effTrojans}.  The $50\%$ detection efficiency for most
of the fields is taken as our limiting magnitude, found to be
$m_{R}=25.7$.  Radii ($r$) of the Neptune Trojans were determined
assuming an albedo of $\rho_{R}=0.05$ and using the equation, $r =
(2.25\times 10^{16} R^{2} \Delta ^{2} / p_{R}\phi (0))^{1/2}
10^{0.2(m_{\odot} - m_{R})}$ where $R$ is the heliocentric distance in
AU, $\Delta$ is the geocentric distance in AU, $m_{\odot}$ is the
apparent red magnitude of the sun ($-27.1$), $p_{R}$ is the red
geometric albedo, $m_{R}$ is the apparent red magnitude of the Trojan
and $\phi (0) = 1$ is the phase function at opposition.  Using an
albedo of 0.05 (Fernandez et al. 2003; Fernandez et al. 2009) we find
that 25.7 magnitudes corresponds to a Neptune Trojan with a radius of
about 16 km.

\begin{figure}
\epsscale{0.4}
\centerline{\includegraphics[angle=0,width=115mm]{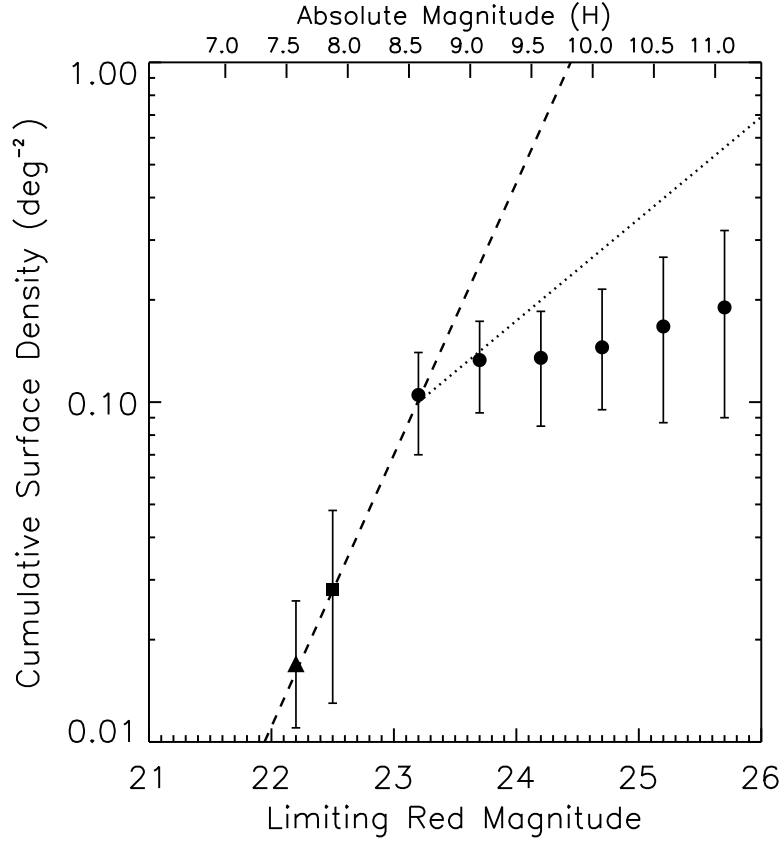}}
\caption{The Cumulative Luminosity Function (CLF) of the Neptune
Trojans, where the black circles are this work, the square is from the
Deep Ecliptic Survey (DES: Chiang et al. 2003) and the triangle is
from the Sloan Digital Sky Survey (SDSS: Becker et al. 2008).  A steep
power law slope ($\alpha = 0.8\pm0.2$ ($q=5\pm1$)), which is similar
to what has been found for the largest objects in the main asteroid
belt, Jupiter Trojans and Kuiper Belt, is plotted as a dashed line and
fits the bright end of the Neptune Trojans CLF.  A roll-over in the
Neptune Trojan CLF is apparent around $m_{R}=23.5\pm0.3$.  The dotted
line shows a shallow power law slope of $\alpha = 0.3$ ($q=2.5$) found
for the intermediate to smaller sized KBOs (Fraser \& Kavelaars 2008;
Fuentes \& Holman 2008).  The data at the faintest end for the Neptune
Trojans are within about $2\sigma$ to the shallow slope found for
faint KBOs.  No Neptune Trojans were discovered fainter than 23.7
magnitudes.  The black circles do not follow a flat horizontal line
when fainter than about 24th magnitude since a smaller area was
covered at fainter magnitudes due to variable seeing conditions on a
few nights.}
\label{fig:cumTrojans}
\end{figure}

The Cumulative Luminosity Function (CLF) describes the sky-plane
number density of objects brighter than a given magnitude.  The CLF
can be described by
\begin{equation}
\mbox{log}[\Sigma (m_{R})]=\alpha (m_{R}-m_{o})  \label{eq:slope}
\end{equation}
\noindent where $\Sigma (m_{R})$ is the number of objects brighter
than $m_{R}$, $m_{o}$ is the magnitude zero point, and $\alpha$
describes the slope of the luminosity function.  The CLF found for
Neptune Trojans is shown in Figure~\ref{fig:cumTrojans}.  The CLF of
the brightest Neptune Trojans ($m_{R}<23.5$ magnitude) follows a steep
power law of $\alpha \sim 0.8$ similar to the brightest Kuiper Belt
objects, Jupiter Trojans and main belt asteroids (Jewitt et al. 2000;
Jedicke et al. 2002; Bottke et al. 2005; Fraser \& Kavelaars 2008;
Fuentes \& Holman 2008).  The Neptune Trojans discovered in our survey
are all bright ($m_{R}<23.7$ magnitudes) compared to the limiting
magnitude of many of the survey fields ($m_{R}=25.7$ magnitudes).  A
roll-over in the Neptune Trojan CLF is apparent around
$m_{R}=23.5\pm0.3$.  For the Neptune Trojans the best fit to the CLF
for $m_{R} < 23.5$ mag is $\alpha = 0.8 \pm 0.2$ and $m_{o} = 24.45
\pm 0.4$.  

\begin{figure}
\epsscale{0.4}
\centerline{\includegraphics[angle=0,width=\textwidth]{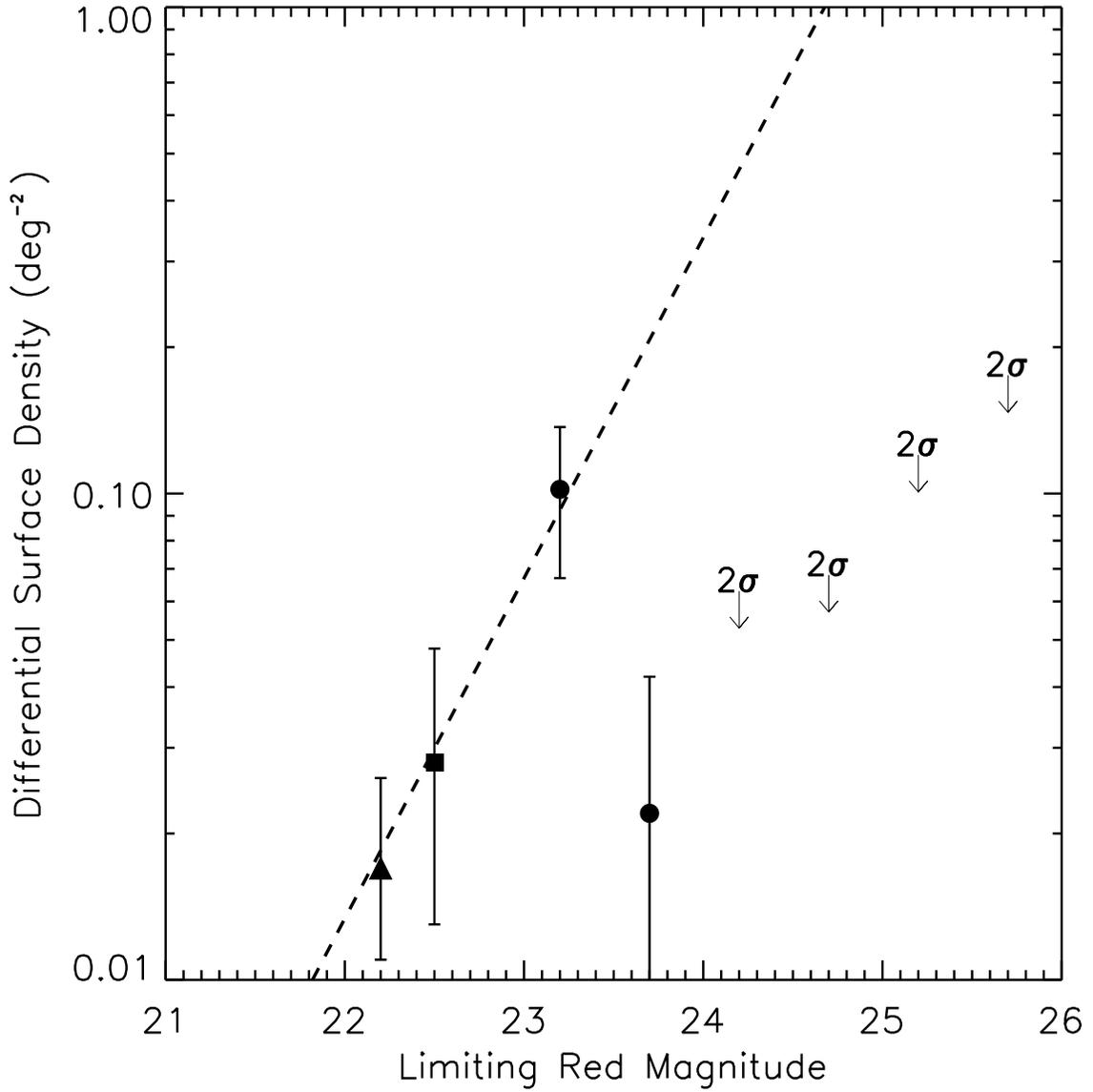}}
\caption{The Differential Luminosity Function (DLF) of the Neptune
Trojans where the black circles are this work, the square is from the
Deep Ecliptic Survey (DES: Chiang et al. 2003) and the triangle is
from the Sloan Digital Sky Survey (SDSS: Becker et al. 2008).  The $2
\sigma$ upper limits are shown for non-detections at fainter
magnitudes where no objects were found in this work.  It is apparent
that there are fewer fainter objects than brighter objects.}
\label{fig:diffTrojans}
\end{figure}

The points in a CLF are heavily correlated with one another, tending
to give excess weight to the faint end of the distribution.  The
Differential Luminosity Function (DLF) does not suffer from this
problem.  We plot the DLF using a bin size of 0.5 mag for all Neptune
Trojans detected in our survey (Figure~\ref{fig:diffTrojans}).  The
roll-over in the number of fainter objects is shown more dramatically
through the DLF, and is insensitive to bin size choice.  If the
fainter (smaller) objects continued to follow the $q=5$ size
distribution slope of the brighter (larger) objects we would have
expected to discover $80\pm 10$ Neptune Trojans between the roll-over
point at $m_{R}\sim 23.5$ magnitudes ($r\sim 45$ km) and our survey
limit of $m_{R}\sim 25.7$ magnitudes ($r\sim 16 $ km).  Though we
found hundreds of Kuiper Belt objects with $m_{R} > 24$ magnitudes we
found zero Neptune Trojans of this faintness, which gives about an
$8\sigma$ result that the smaller Neptune Trojans have a shallower
power-law slope than the larger Neptune Trojans.

Thus like the other known stable small body reservoirs, the CLF of the
Neptune Trojans is best fit by a broken power law that breaks from a
steep slope for the largest objects to a shallow distribution for the
smaller objects.  The data at the faintest end for the Neptune Trojans
are within about $2\sigma$ of the shallow slope found for faint KBOs
(Figure~\ref{fig:cumTrojans}).  Further data are required to determine
a reliable slope for the Neptune Trojans at the faint end of the CLF.
Though we cannot completely rule out the possibility that the Neptune
Trojans have a single very shallow power law it seems unlikely since
we cannot fit a single power law to all the points to within $1\sigma$
and the best fit power law to all the data would be the shallowest
observed ($q\sim 2.5$) for such relatively large objects in the solar
system.  This slope would also be much shallower than the Dohnanyi
slope of 3.5 that is expected if the objects were in a state of
collisional equilibrium (O'Brien \& Greenberg 2003).

\section{Size Distribution}

\begin{figure}
\epsscale{0.4}
\centerline{\includegraphics[angle=0,width=\textwidth]{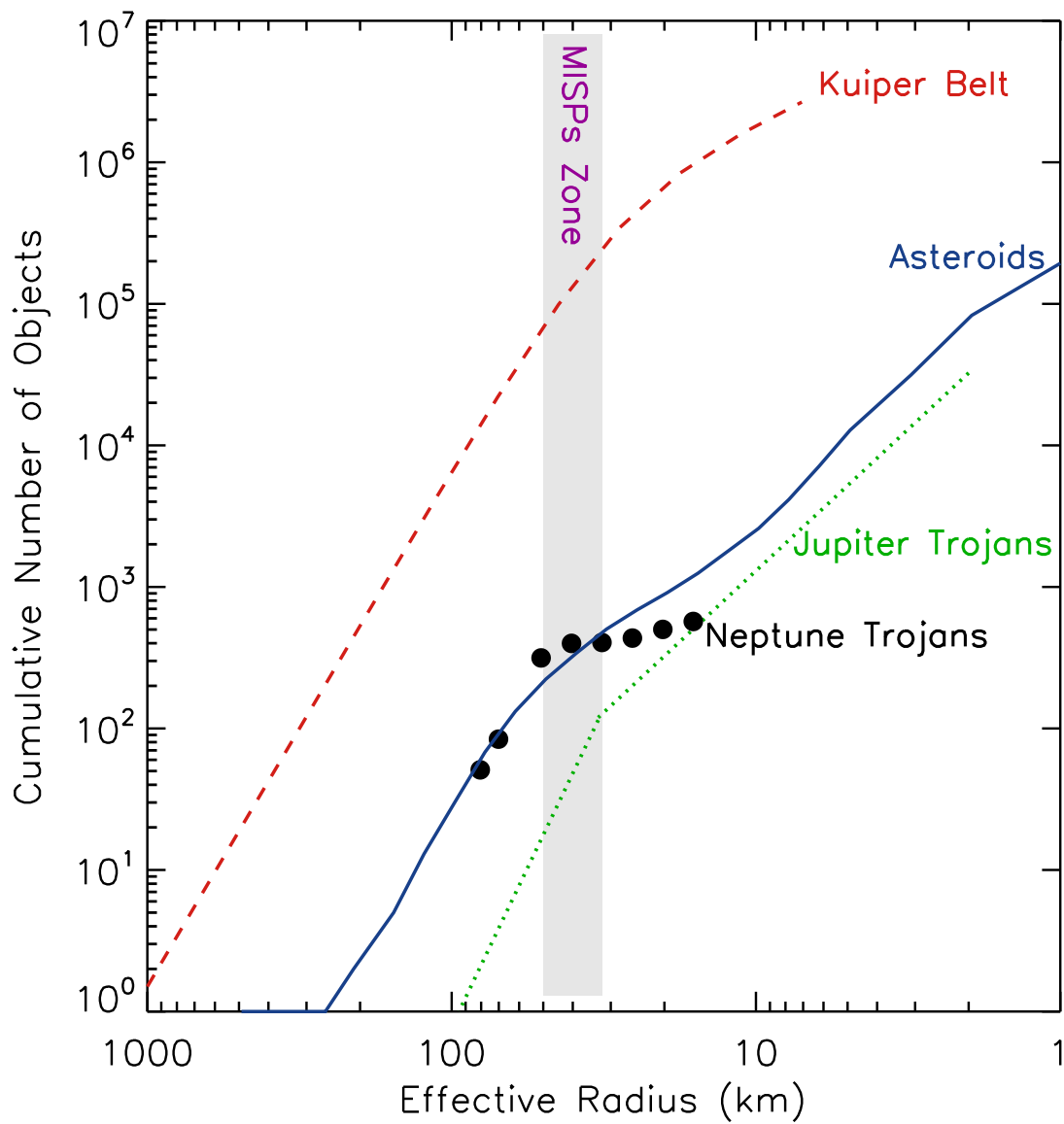}}
\caption{The cumulative size distribution of the Neptune Trojans
(black circles; from the SDSS, DES and this work which includes both
the L4 and L5 clouds), Jupiter Trojans (Jewitt et al. 2000) (green
dotted line; including both the L4 and L5 clouds), Kuiper Belt objects
(Fraser and Kavelaars 2008; Fuentes and Holman 2008) (red dashed line)
and main belt asteroids (Jedicke et al. 2002; Bottke et al. 2005)
(blue solid line).  All four small body reservoirs have a similar
steep slope for the largest objects size distribution ($q\sim5$).
Though the objects in each reservoir likely have significantly
different compositions, internal structures and collisional histories
they all have a roll-over in their size distributions between about 35
and 50 km in radius, which we call the Missing Intermediate Sized
Planetesimals (MISPs).  The similar size of the roll-over for each
stable reservoir favors a primordial instead of a collisional
formation.  The error bars for the Neptune Trojan points have been
removed for clarity but would be similar as in Figure 2.}
\label{fig:sizeTrojans} 
\end{figure}

One of the main ways to constrain the formation and collisional
history of an ensemble of objects such as the Neptune Trojans is to
determine their size distribution.  The size distribution is an
indicator of how the accretion process worked and is related to the
CLF.  If we assume the Neptune Trojans all have similar albedos to the
Jupiter Trojans (0.05, Fernandez et al. 2003) and are at a distance of
30 AU we can determine the size distribution
(Figure~\ref{fig:sizeTrojans}) of the Neptune Trojans from the CLF.  A
slope of $q=5\pm1$ is found for the large Neptune Trojan size
distribution, where $n(r)dr \propto r^{-q}dr$ is the differential
power-law radius distribution with $n(r)dr$ the number of Neptune
Trojans with radii in the range $r$ to $r+dr$.

The roll-over in the size distribution for the Neptune Trojans occurs
around a radius of $45\pm10$ km.  When compared to the Kuiper Belt
(Bernstein et al. 2004; Fraser \& Kavelaars 2008; Fuentes \& Holman
2008; Schlichting et al. 2009) and Jupiter Trojan (Jewitt et al. 2000;
Yoshida \& Nakamura 2005; Yoshida \& Nakamura 2008) size distributions
it appears that all three outer solar system small body populations
have a similar roll-over in their size distribution (radii of
$35\pm5$, $40\pm15$ km respectively for the Jupiter Trojans and Kuiper
Belt).  Previous authors have shown that the Kuiper Belt's low
inclination and high inclination classical populations may have
different initial size distributions (Bernstein et al. 2004; Fuentes
\& Kavelaars 2008; Morbidelli et al. 2009).  The Neptune Trojans have
a steep size distribution at the large end, similar to that found for
the low inclination Kuiper Belt objects and Jupiter Trojans, which may
hint at a common origin (Morbidelli et al. 2009).  Further, the main
belt of asteroids (Jedicke et al. 2002; Bottke et al. 2005) has a
similar roll-over as the above reservoirs around a radius of $50\pm5$
km.

\subsection{Missing Intermediate Sized Planetesimals (MISPs)}

There are unexpectedly far fewer tens of kilometer sized objects than
larger objects in all the known stable reservoirs of small bodies in
our solar system.  The Missing Intermediate Sized Planetesimals
(MISPs) are either an indication that the intermediate sized objects
have been collisionally eroded over the age of the solar system or a
primordial distribution deficient in intermediate sized objects was in
place before capture.  MISPs are not expected from a uniformly
accreted population (Bottke et al. 2005; Kenyon et al. 2008).
Detailed numerical collisional simulations (Morbidelli et al. 2009)
have shown that collisional grinding is not the likely cause of the
roll-over around a radius of 50 km in the size distribution of the
main belt asteroids.  This suggests the roll-over is a consequence of
the original primordial formation mechanism of planetesimals.
Further, considering that each of the different small body reservoirs
have significantly different compositions and collisional histories,
the similarities of all the MISPs sizes throughout the various
reservoirs suggests the lack of intermediate sized objects is likely a
primordial result from before the planets formed.  This has
far-reaching implications for planetesimal formation and is very
important for understanding planet formation in general.  It is
currently not understood how objects larger than about a meter in size
formed in the solar nebula since meter-sized objects are likely to be
disrupted from collisional erosion and have very short dynamical
lifetimes due to gas drag and thus should not be able to grow to
larger sizes (Wurm et al. 2005; Cuzzi \& Weidenschilling 2006; Dominik
et al. 2007; Blum et al. 2008). Several recent papers have suggested
that the planetesimal formation mechanism jumps over this meter size
barrier and instead large objects of hundreds of kilometers in size
coalesce directly from over-dense clouds of cm to meter sized
particles in a highly turbulent primordial solar nebula (Johansen et
al. 2007; Cuzzi et al. 2008; Morbidelli et al. 2009; Johansen et
al. 2009).

Numerical collisional simulations as performed for the main asteroid
belt (Morbidelli et al. 2009) and their effect on the size
distribution for the outer solar system small body reservoirs are
warranted to constrain the role collisions play in helping produce the
observed size distributions in these locations.  It is likely that the
Trojans and Kuiper Belt objects have undergone less collisional
evolution than the main belt asteroids since emplaced in their current
orbits (Kenyon et al. 2008; Morbidelli et al. 2009; Levison et
al. 2009) and if confirmed would indicate the observed roll-over in
their size distribution is likely from primordial formation of the
planetesimals.  In addition, the outer Solar System objects are likely
to have some material strength (Levison et al. 2009; Leinhardt \&
Stewart 2009), which appears to be an impediment for collisional
erosion to account for the observed roll-over in the Kuiper Belt size
distribution (Pan and Sari 2005).  To date the size distributions of
these more distant outer solar system reservoirs are much more poorly
characterized by observations as the main asteroid belt.

Detailed in this work is the first measurement of the size
distribution for the Neptune Trojans.  The observations in this work,
when compared with limited collisional simulations (Morbidelli et
al. 2009), show that the lack of objects tens of kilometers in size
for all known reservoirs agrees with planetesimal formation skipping
over forming significant numbers of objects in the tens of kilometer
size range for all areas of the Solar System.  In this scenario,
objects smaller than about $r \sim 35-50$ km are most likely
collisional by-products of larger primordial objects (Farinella \&
Davis 1996; Kenyon \& Bromley 2004; Kenyon et al. 2008).  It is
possible that future collisional simulations using size-dependent
drift due to the drag of a low turbulent solar nebular gas during
accretion could account for the observed roll-over (Weidenschilling
2010) in the size distributions, but this scenario would likely be
much more dependent on formation location in the solar nebula unlike
the primordial coalescence of large objects from over-dense dust
clouds.

\subsection{Small Body Reservoir Populations}

Comparing the various observed small body reservoirs shows that the
Kuiper Belt holds most of the objects larger than 50 km in radius,
about $150,000$.  The Neptune Trojans are the next most populated
reservoir with about 400 objects larger than 50 km in radius expected
(a factor of about 375 less than the Kuiper Belt).  The main belt
asteroids contain about 200 objects larger than 50 km in radius or a
factor of 2 less than the Neptune Trojans.  The Jupiter Trojans have
about 50 objects this size or a factor of 8 less than the Neptune
Trojans.

\section*{Acknowledgments}

We thank J. Chambers for comments on this manuscript.  Based in
part on data collected at Subaru Telescope, which is operated by the
National Astronomical Observatory of Japan.  This paper includes data
gathered with the 6.5 meter Magellan Telescopes located at Las
Campanas Observatory, Chile.  C.T. was supported by the Gemini
Observatory, which is operated by the Association of Universities for
Research in Astronomy, Inc., on behalf of the international Gemini
partnership of Argentina, Australia, Brazil, Canada, Chile, the United
Kingdom, and the United States of America. S. S. was supported in part
by funds from the NASA New Horizons Spacecraft Pluto mission.


\newpage

\newpage

\newpage

\newpage

\end{document}